\documentclass[useAMS,usenatbib]{mn2e}
\usepackage{graphicx}
\usepackage[figuresright]{rotating}
\usepackage{subfigure}
\usepackage{amssymb}

\newcommand{\ii}{\'\i}
\newcommand{\ion}[2]{#1 {\sc #2}} 

\usepackage{color}
                                             
\title[A study of the interacting system AM\,1219--430]
{Photometry and Dynamics of the Minor Merger AM\,1219-430 with Gemini\/ GMOS-S}

\author[Hernandez-Jimenez et al.]
{J.~A.~Hernandez-Jimenez$^1$\thanks{E-mail:hernandez.jimenez@ufrgs.br}, 
M.~G.~Pastoriza$^1$,  I.~Rodrigues$^2$, A.~C.~Krabbe$^2$, \and Cl\'audia ~Winge$^3$  C.~Bonatto$^1$ \\
$^1$ Instituto de F\ii sica, Universidade Federal do Rio Grande do Sul, Av.~Bento Gon\c{c}alves,9500, Cep 91501-970, Porto Alegre, RS, Brazil\\
$^2$ Universidade do Vale do Para\'iba, Av. Shishima Hifumi, 2911, Cep 12244-000, S\~ao Jos\'e dos Campos, SP, Brazil\\
$^3$ Gemini Observatory, c/o AURA Inc., Casilla 603, La Serena, Chile\\
}
 
\begin{document}

\date{Accepted -. Received -.}

\pagerange{\pageref{firstpage}--\pageref{lastpage}} \pubyear{2006}

\maketitle

\label{firstpage}

\begin{abstract}

We present an observational study of the interaction effect on the dynamics and morphology of the minor 
merger AM\,1219-430. This work is based on  $r'$ and $g'$ images and  long-slit spectra obtained with the 
Gemini Multi-Object Spectrograph at the Gemini South Telescope. We detected  a tidal tail in the
main galaxy (AM\,1219A) and a bridge of material connecting the galaxies. In luminosity, AM\,1219A 
is about 3.8 times brighter than the secondary (AM\,1219B). The surface brightness profile of AM\,1219A was
decomposed into bulge and disc components. The profile shows a light excess of $\sim 53 \%$ 
due to the contribution of star-forming regions, which is typical of starburst galaxies. On the 
other hand, the surface brightness profile of AM\,1219B shows a lens structure in addition to the 
bulge and disc. The scale lengths  and central magnitudes of the disc structure of both galaxies agree with the
average values derived for galaxies with no sign of ongoing interaction or disturbed morphology. The S\'ersic 
index ($n<2$), the effective and scale radii of the bulge of both galaxies are typical of pseudo-bulges. The rotation 
curve of AM\,1219A derived from the emission line of ionized gas is quite
asymmetric, suggesting a gas perturbed by interaction.  We explore all
possible values of stellar and dark matter masses. The overall
best-fitting solution for the mass distribution of AM\,1219A was found with M/L for bulge and disc of
$\Upsilon_{b}=2.8_{-0.4}^{+0.4}$ and $\Upsilon_{d}=2.4_{-0.2}^{+0.3}$, respectively, and a \citeauthor{navarro95} profile  of $M_{200}=2.0_{-0.4}^{+0.5}\times10^{12}\, M_{\odot}$ and $c=16.0_{-1.1}^{+1.2}$. The estimated 
dynamical mass  is $1.6\times10^{11}\, M_{\odot}$, within a radius of $\sim10.6$ kpc.

\end{abstract}


\section{Introduction}

Since the pioneering works of  \citet{vorontsov59} and \citet{arp66}  classifying hundreds  of ``peculiar galaxies'', 
now known as interacting and merger galaxies, the number of studies on the nature of these objects has grown exponentially. 
Now it is clear that interactions and merger events represent important mechanisms for 
driving the evolution of galaxies (see review of \citealt{struck06}).  They can lead to the formation of morphological 
structure, such as tidal tails like those detected in Superantennae (\citealt{mirabel91}, see the review of 
\citealt{duc13});  bridges of stars and gas like that observed in M\,51 \citep{lee12}. In addition, the velocity fields 
of interacting galaxies show asymmetries and irregularities due to the interaction with 
the companion galaxy \citep[e.g.,][]{rubin91,rubin99,dale01,mendes03,fuentes04,presotto10}. 
The kinematic disturbances are expected to disappear in about 1 Gyr after the first encounter \citep{kronberger06}. One signature of 
perturbation is the so called ``bifurcation'', when the rotation curve in one side of the disc declines while the other side 
remains steady \citep{pedrosa08}.

Interaction events can be divided into major and minor mergers. In major mergers, 
the masses of the involved galaxies are comparable. On the other hand, minor mergers are 
those in which a large galaxy  interacts with a dwarf or low-mass galaxy. Although major 
mergers are a striking phenomena  and thus  have received most of the attention \citep{schwarzkopf00}, 
they are less common than minor mergers. In fact,
minor mergers of galaxies occur at least an order of magnitude more  
often  than major mergers \citep{hernquist95}. In addition, hierarchical models of cosmological 
structure formation predict that galaxies grow  by accreting other galaxies, 
more frequently minor companions  \citep[e.g.,][]{cole00,wechsler02,bedorf12}.

Obtaining photometric and kinematic information on minor merger systems is useful for understanding
the effect that interaction may have on each component. The decomposition of  the surface brightness profile 
into bulge and disc components \citep{fathi10a}, for instance, allows us to infer the stellar mass distribution. 
This information, together with the rotation curve, is used to constrain models of dark matter 
distribution \citep{vanalbada85,carignan85,kent87,blais01}. 
These parameters are important to test the predictions of minor merger simulations and thus, 
to reconstruct their dynamical history (e.g., \citealt{salo93,mihos97,irapa1999,irapa2000,thies01,krabbe08,krabbe11}, 
and see  \citealt{barnes09} for a partial list of system modelled). 
 
\begin{figure*}
\centering
\includegraphics*[width=\textwidth]{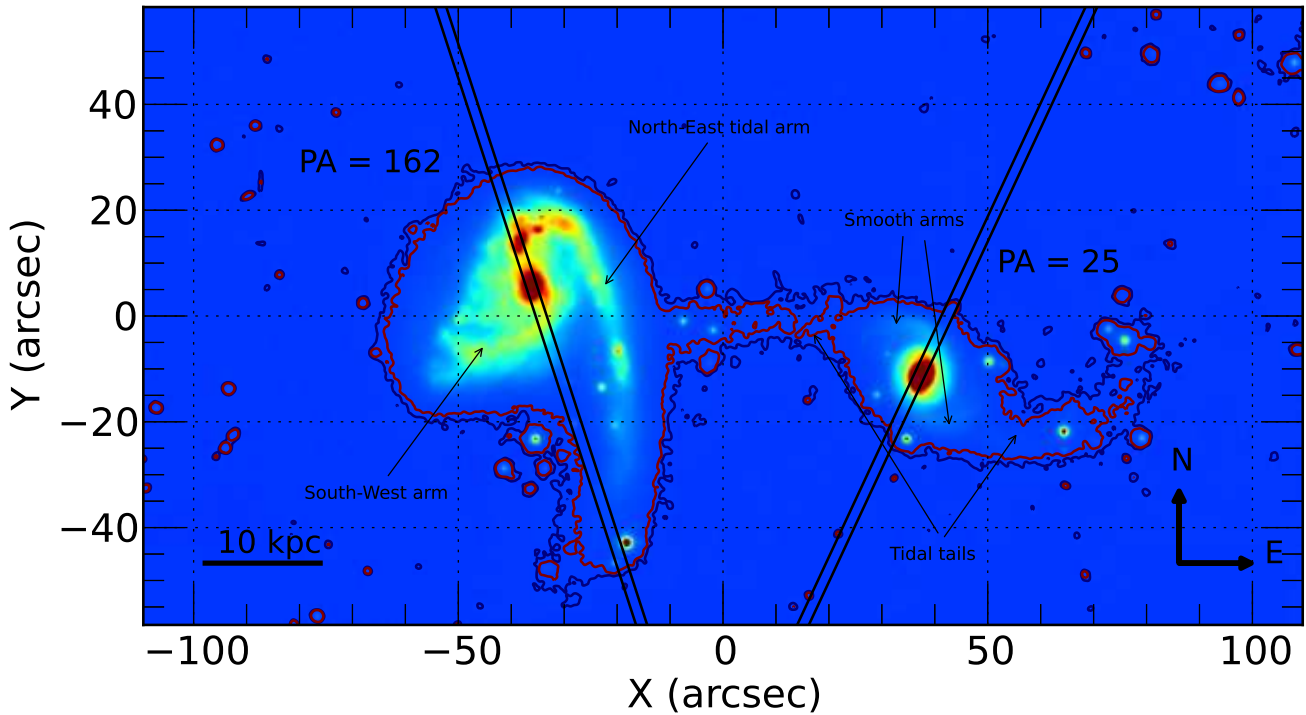}  				     
\caption{AM\,1219-430 $r'$ image with the observed slit positions. The striking bridge of material between the
galaxies and tidal arms shows up by isophotes with 2$\sigma$ (blue) and 3$\sigma$ (red) above the
sky value.}
\label{contours}
\end{figure*}

In order to understand the effect of the interaction on the kinematic and photometric properties of 
minor merger components, we have conducted  long-slit spectroscopy and $g'$ and $r'$ images observations 
with Gemini Multi-Object Spectrograph (GMOS) at Gemini South Telescope. The pair systems were 
selected from  \citeauthor{donzelli97}'s (\citeyear{donzelli97}) sample. The first results 
of this programme have been presented for AM\,2306--721 \citep{krabbe08} and AM\,2322--821 \citep{krabbe11}. In this work, we 
study the system AM\,1219--430, shown in Fig. \ref{contours}. This pair is among those showing strong star formation activity 
in the above sample. In fact, stellar population analysis based on equivalent widths of absorption lines and continuum distribution 
shows that  both components  have a strong flux contribution from stellar populations younger than $10^8$ years \citep{pastoriza99}. 
Besides, the  emission line ratios of both components  are typical  of \ion{H}{ii} region spectra \citep{pastoriza99}. 
In addition, \citet{kewley01} calculated the infrared luminosity ($L_{IR}$) of AM\,1219A and found that it is a luminous 
infrared galaxy with $L_{IR}=1.26 \times 10^{11}\,L_{\sun}$.

The main morphological features of the primary galaxy (hereafter, AM\,1219A) are  (see labels in Fig. \ref{contours}): a 
normal arm  curled  to the south-west and a strong tidal arm curled to the north-east, with several star formation 
knots along it. There is a giant \ion{H}{ii} region at the nucleus and at the north, where the tidal arm turns
to the south. On the other hand, the secondary galaxy (AM\,1219B) has a bright bulge 
with smooth open spiral arms, and symmetric tidal tails. The west tail seems to form a bridge between the galaxies.

This paper is organized as follows: in Section \ref{datared}, we provide details about the observations and data reduction, 
photometric calibrations, measurement of the sky background and integrated magnitudes of the galaxies. Section \ref{restima} 
explains the image restoration, symmetrization method to separate the symmetric and non-symmetric structures in the galaxies, 
and photometric decomposition of the surface brightness profiles. Section \ref{vel} describes the gas kinematics and the 
determination of the rotation curve. In Section \ref{massmodel}, we present the models of bulge, disc and halo components 
used to decompose the rotation curve. In Section \ref{fitmodel}, we discuss the methods to fit the rotation curve and the 
determination of the mass-to-light ratio (M/L) of each component and halo parameters.  Finally, the conclusions are 
summarized in Section \ref{final}. Throughout this paper, we adopt the Hubble constant as $H_0$=73\,km\,s$^{-1}$\,Mpc$^{-1}$ \citep{spergel07}.

\section{Observations and data reduction}
\label{datared}

This paper is based on $g'$ and $r'$ images and  long-slit spectra  obtained in 2007:  
February 12 and May 23 with the GMOS at Gemini South 
Telescope, as part of the poor weather programme GS-2007A-Q-76. We obtained science
quality (slow readout) sets of short exposure time $r'$ and $g'$ images. The  images were  binned by 2 pixels, 
resulting in a spatial scale of 0.146 arcsec pixel$^{-1}$. They were  processed using
standard procedures (bias subtraction and flat-fielding) and combined to obtain the final $g'$ 
and $r'$ images. Seeings of $\sim0.99$ and $\sim1.26$ arcsec were estimated for  $g'$ and $r'$, respectively.
The seeing was calculated by fitting a Gaussian  profile to the field star images. The journal of observations is presented in Table \ref{observ_ima}.

\begin{table}
\caption{Journal of image observations }
\label{observ_ima}
\begin{tabular}{lccc}
\noalign{\smallskip}
\hline
\noalign{\smallskip}
Date (UT) & Exposure time (s) & Filter & $\Delta \lambda$ (\AA) \\
\noalign{\smallskip}
\hline
\noalign{\smallskip}
2007--12--02  &  5$\times$30    & $g^{\prime}$ (G0325)       & 3980--5520  \\
2007--12--02  &  4$\times$30    & $r^{\prime}$ (G0326)      & 4562--6980   \\
\noalign{\smallskip}
\hline
\noalign{\smallskip}
\end{tabular}
\end{table}

Spectra  in the range 4\,280 - 7\,130\AA\ were obtained  with the B600 grating,
plus  the 1 arcsec slit, which gives a spectral resolution of 5.5\AA. 
The frames were binned on-chip by 4 and 2 pixels in the spatial and wavelength directions, respectively, 
resulting in a spatial scale of 0.288 arcsec pixel$^{-1}$, and 0.9\,\AA\,pixel$^{-1}$ dispersion. 

Spectra at two different position angles (PA) were taken:  PA=162$\degr$ and 25$\degr$ , which correspond to the position 
along the apparent major axis of AM\,1219A and the minor axis of AM\,1219B, respectively.
The exposure time of each single frame was limited to 700 s to 
minimize the effects of cosmic rays, with four frames being obtained
for each slit position to achieve suitable signal-to-noise ratio. The slit positions are shown in Fig. \ref{contours}, 
superimposed on the $r'$-band image of the pair. Table \ref{observ_spec}
gives the journal of spectroscopic observations.

\begin{table}
\caption{Journal of long-slit observations}
\label{observ_spec}
\begin{tabular}{lccc}
\noalign{\smallskip}
\hline
\noalign{\smallskip}
Date (UT) & Exposure time (s) & PA (\degr)& $\Delta \lambda$ (\AA) \\
\noalign{\smallskip}
\hline
\noalign{\smallskip}
2007--05--23  &  4$\times$700    & 162      & 4280--7130  \\
2007--05--23  &  4$\times$700    & 25       & 4280--7130   \\
\noalign{\smallskip}
\hline
\noalign{\smallskip}
\end{tabular}
\end{table}

The spectroscopic data reduction was carried out  using 
the {\sc gemini.gmos} package as well as  generic {\sc IRAF}\footnote{IRAF is distributed by 
the National Optical Astronomy Observatories, which is operated by the Association of
Universities for Research in Astronomy, Inc. (AURA) under cooperative agreement with 
the National Science Foundation.} tasks. We followed the standard procedure for
bias correction, flat-fielding, cosmic ray cleaning, sky subtraction,
wavelength, and relative flux calibrations. In order to increase the
signal-to-noise ratio, the spectra were extracted by summing over
four rows. Each spectrum thus represents an aperture of 1 arcsec $\times$
1.17 arcsec. We adopted 95\,Mpc as distance of the pair (estimated from the radial 
velocity of the central aperture of AM\,1219A, see  Section  \ref{vel}); thus, 
this aperture corresponds to a region of 460\,$\times$\,539\,pc$^2$.

\begin{figure}
\centering
\includegraphics*[width=\columnwidth]{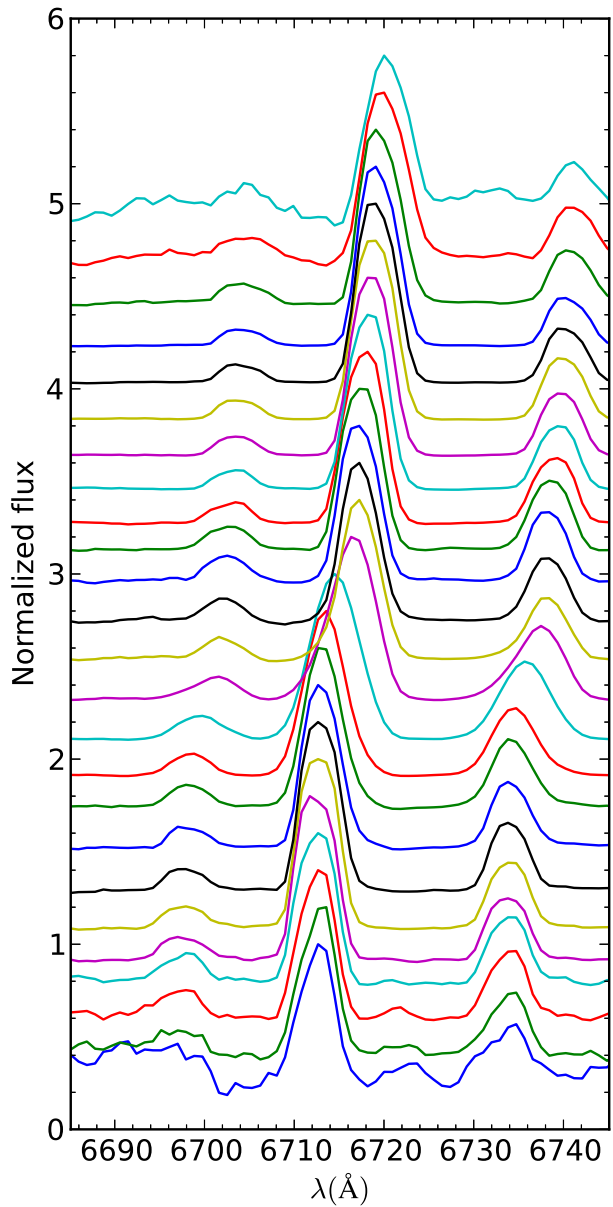}  				     
\caption{Normalized spectra  showing the lines $\rm H\alpha$ and [\ion{N}{ii}] from 
long-slit apertures along the apparent major axis of AM\,1219A.} 
\label{spectraa}
\end{figure}

The nominal centre of each slit corresponds to  the continuum peak at $\lambda\,5525$\AA. 
Fig.~\ref{spectraa} shows a sample of  spectra of AM\,1219A  extracted along  the  slit positions at PA=162$\degr$.
The spectra were normalized and a constant  was added in order to illustrate  the 
main features of the observed regions as well as the receding 
and approaching galaxy sides.   

\subsection{Photometric calibration}
\label{calima}

In order to calibrate the $r'$ and $g'$ images, we used  seven foreground stars present in
the GMOS field, which are in the United States Naval Observatory-B1.0 Catalogue \citep[USNO-B;][]{monet03}.
Aperture photometry of these stars was carried out using the {\sc PHOT} task within 
{\sc IRAF/DAOPHOT}. The aperture radii of 3.7 arcsec (25 pixel)
and 4.4 arcsec (30 pixel) were used for the $g'$ and $r'$ images, respectively. 
We applied the bandpass transformation given by \citet{monet03} to convert the J and F photographic magnitudes 
to the $g'$ and  $r'$ magnitudes in the SDSS system. Then, the zero-points for the $g'$ and $r'$ images were found according to:

\begin{equation}
g'=28.52\pm 0.31 -2.5\log(C/t)
\label{rprima}
\end{equation}

\begin{equation}
r'=28.29\pm 0.24 -2.5\log(C/t),
\label{gprima}
\end{equation}
where $C$ is the integrated counts within 3.7 and 4.4 arcsec  aperture
radii in $g'$ and $r'$ images, respectively, and $t$ is the exposure time.

\subsection{Sky background}
\label{skyback}

The sky background levels of the $g'$ and $r'$ images were adopted as the mean value of 
several boxes of $60\times60$ pixels. The selected regions are far from stars and galaxies, so that the 
mean value is not biased by any residual emission. The statistical
standard deviation ($\sigma$) of the sky background around the mean value was also computed for
these regions. $\sigma$ is an estimate of the sky noise, and
the value of 1 $\sigma$ determines the physical limit for galaxy extension. 
Table \ref{sigma} shows the detection limits, in magnitudes per
square arcsecond, of the $g'$ and $r'$ images measured at 1, 2, and 3 $\sigma$. 

The striking bridge of material connecting the galaxies and tidal arms was detected by drawing
isophotes with levels corresponding to 2$\sigma$  and 3$\sigma$ over the sky background (Fig. \ref{contours}).

\begin{table}
\caption{Sky background levels }
\label{sigma}
\begin{tabular}{lcccc}
\noalign{\smallskip}
\hline
\noalign{\smallskip}
Filter & 1$\sigma$  & 2$\sigma$ & 3$\sigma$  \\
\noalign{\smallskip}
\hline
\noalign{\smallskip}
   $g'$    & 26.09   & 25.34  & 24.89 \\
   $r'$    & 26.05   & 25.30  & 24.86 \\

\noalign{\smallskip}
\hline
\noalign{\smallskip}
\end{tabular}
\end{table}

Table \ref{magpar} gives the integrated magnitudes $m_{r'}$ and $m_{g'}$ as well as the absolute 
magnitudes and luminosities of both galaxies. Magnitudes were calculated by integrating the flux 
inside the isophote with an intensity brighter than that of the tidal arms and the bridge between 
galaxies. The surface brightness of these isophotes ($\mu_{lim}$) is also given in Table \ref{magpar}. 
The absolute magnitudes were corrected for the Galactic extinction, adopting  
$A_{g}=0.25$ and $A_{r}=0.36$ \citep{schlafly11}. Finally, the  luminosities  were estimated from 
the solar absolute magnitudes in $g'$ (5.45) and $r'$ (4.76) bands \citep{blanton03} . These values, 
in turn were used to derive the integrated and absolute magnitudes, and luminosity  for the $B$ band 
in the Jonhson system, using the  \citeauthor{fukugita96} (\citeyear{fukugita96})   
transformations (adopting $A_{B}$=0.397 and $M_{\odot}(B)=5.48$). These values are also 
listed in Table \ref{magpar}, and they will be useful to calculate the M/L in 
Section \ref{massmodel}.

\begin{table}
\caption{Total magnitudes and luminosities}
\label{magpar}
\begin{tabular}{lccccccc}
\noalign{\smallskip}
\hline
\noalign{\smallskip}
Galaxy & Filter &  $m_{\scriptsize{\mbox{T}}}$ & $M_{\scriptsize{\mbox{T}}}$ &  $L/\mbox{L}_{\odot}$ & $\mu_{lim}$ ($mag/arcsec^{2}$) \\
   & [1] & [2] & [3] & [4] & [5] \\
\noalign{\smallskip}
\hline
\noalign{\smallskip}
AM\,1219A  & $g^{\prime}$ & 14.48 & $-20.76$  & $3.07\times10^{10}$ &25.31 \\
           & $r^{\prime}$ & 13.43 & $-21.71$  & $3.87\times10^{10}$ &23.78 \\
           & $B$          & 15.13 & $-20.15$  & $1.79\times10^{10}$ &25.17 \\
AM\,1219B  & $g^{\prime}$ & 16.00 & $-19.25$  & $7.57\times10^{9}$  &25.25 \\
           & $r^{\prime}$ & 14.84 & $-20.30$  & $1.05\times10^{10}$ &24.01 \\
           & $B$          & 16.70 & $-18.58$  & $4.23\times10^{9}$  &25.98 \\
\noalign{\smallskip}
\hline
\noalign{\smallskip}
\end{tabular}
\end{table}

\section{Image restoration}
\label{restima}

In order to improve the identification of \ion{H}{II} regions,  we
deconvolved the $g'$ and $r'$ images using  Lucy--Richardson (L-R) algorithm
\citep{richardson72,lucy74}. The first step is to create a synthetic point spread function (PSF) from the calibration stars using  the {\sc PSF} task of {\sc IRAF/DAOPHOT}. We choose  a 2D elliptical Gaussian for the functional form of the analytic component of the PSF model. 
Briefly, the algorithm computes the parameters of the analytic function by  fitting all 
the stars weighted by their signal-to-noise  ratio using a non-linear square fitting 
technique. The reader is referred to \citet{stetson87} for more details. Then, we deconvolved 
the images using the {\sc LUCY} task within {\sc IRAF/STSDAS}. 
 The  L-R algorithm generates a restored (or deconvolved) image  through an
iterative method.  The essence of L-R is as follows: at the end of each 
iteration, the resulting reconstructed image is convolved with the PSF and 
compared to the observed image. A correction image is generated by the 
ratio of the observed image to its PSF-convolved reconstruction, which measures 
the fractional deviation between the model and the observed. This information is
then used to refine the reconstruction process during the next 
iteration. This correction and comparison process is repeated until the 
correction image no longer changes significantly \citep{pogge02}. The restored data will be properly normalized,  
and  the  integrated flux  in  the  image  is conserved.  The total flux for a particular object should also be conserved.
The deconvolved  $r'$ image of AM\,1219A is present in the left-panel of Fig. \ref{elmeA}. This procedure resolved 
the compact nucleus in several \ion{H}{ii} regions. In addition, we can see that the spiral arms start at the very 
centre of the galaxy, and the giant \ion{H}{ii} region of the strong perturbed north-east arm is resolved in \ion{H}{ii} 
region complexes. The deconvolved   $r'$ image of AM\,1219B is present in the 
left-panel of Fig. \ref{elmeB}. This image shows no significant difference from the original one. We also did not 
detected  \ion{H}{ii} region complexes as in the main galaxy.

\subsection{Elmegreen method}
\label{elme}

To verify the level of the morphological distortions introduced by the interaction, we applied the symmetrization method of \citet{eem92} to the galaxies.  This method is based on the $m$-fold symmetry concept:  if the spiral structure remains invariant under a rotation of $2 \pi /m$ around the image centre ($m$ is the number of arms), the galaxy has $m$ dominant arms. In most galaxies, the dominant symmetry is $m=2$. Following this procedure, we have separated both the two-fold symmetric and non-symmetric parts of the spiral galaxy pattern, by making successive image rotations and subtractions. The non-symmetric image (hereafter $A_{2}$)  is obtained by subtracting from the observed image the same image rotated by $\pi$. If $I(r,\theta)$ is the original image in polar coordinates, then

\begin{equation}
A_{2}=[I(r,\theta)-I(r,\theta+\pi)]_{\scriptsize{\mbox{T}}},
\end{equation}
where the subscript  {\scriptsize T} stands for truncation, meaning that pixels with negative intensities are set to 0. 
On the other hand, the symmetric image (hereafter $S_{2}$) is obtained by subtracting the asymmetric image  from the observed one. The  
$S_{2}$ image shows what should be the ``original disc'' and the non-perturbed spiral pattern. Therefore, the symmetric images correspond to

\begin{equation}
S_{2}=I(r,\theta)-A_{2}(r,\theta).
\end{equation}

\begin{figure*}
\centering
\includegraphics*[width=\textwidth]{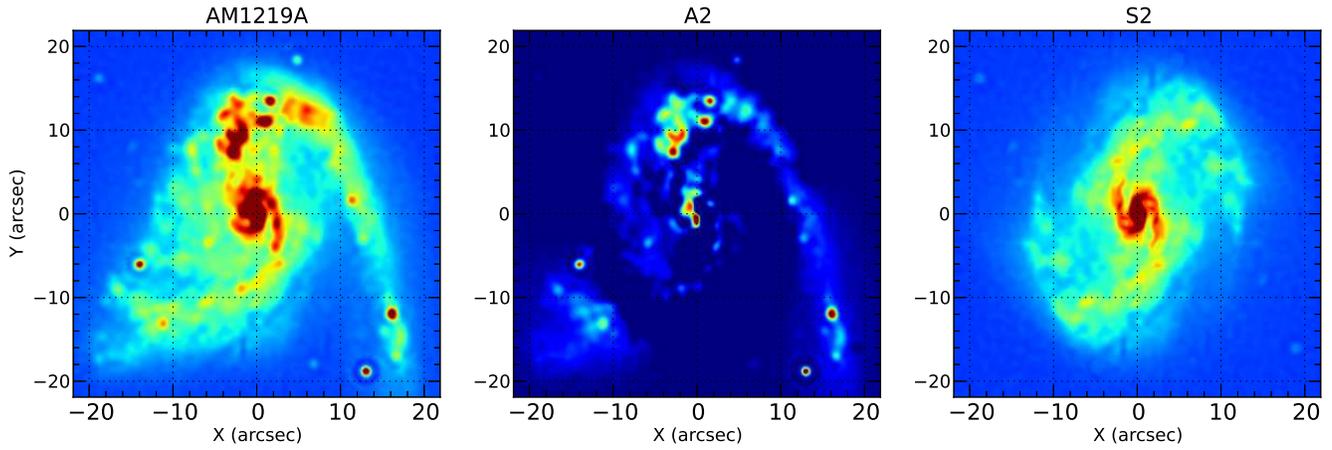}  				     
\caption{Elmegreen method  applied to AM\,1219A. The left-hand panel shows the deconvolved image in the $r'$ band, the middle panel shows the
$A_{2}$  image and the right-hand panel the $S_{2}$ image.}
\label{elmeA}
\end{figure*}

\begin{figure*}
\centering
\includegraphics*[width=\textwidth]{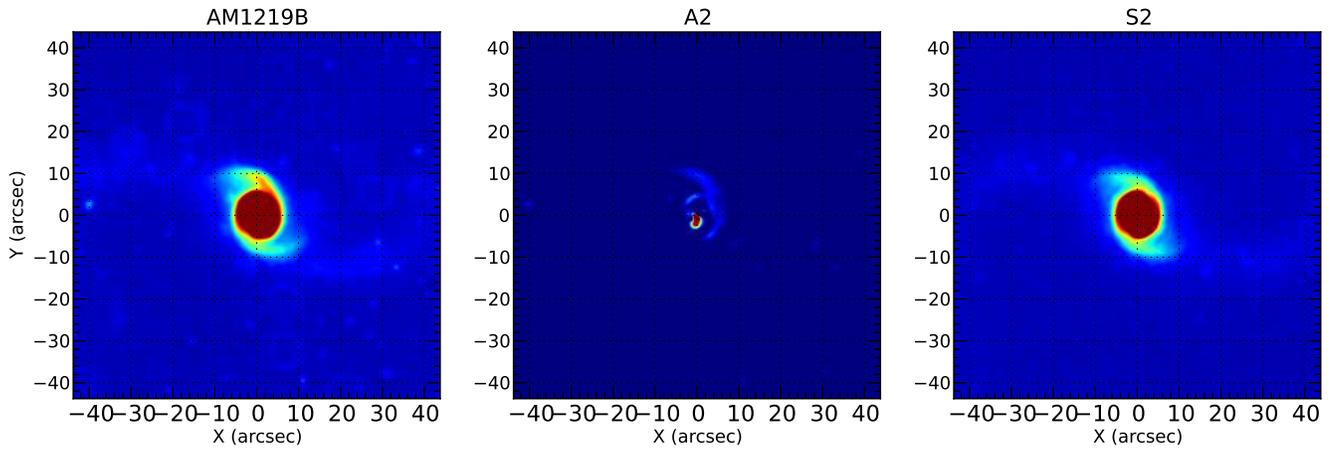}  				     
\caption{Same as Fig. \ref{elmeA} for AM\,1219B.}
\label{elmeB}
\end{figure*}

The results for AM\,1219A and AM\,1219B are displayed in  Figs \ref{elmeA} and \ref{elmeB},
respectively. The left-hand panels show the deconvolved images of the galaxies in $r'$, the middle panels the 
$A_{2}$  images and the right-hand panels the $S_{2}$ images. The  $A_{2}$  image of AM\,1219A shows a one-arm structure, 
with several \ion{H}{ii} region complexes along it. The $S_{2}$  image ``dig up'' the non-perturbed spiral pattern of AM\,1219A, 
from which we can classify the non-perturbed structure of AM\,1219A as Sbc galaxy type. On the other hand, the $A_{2}$ image of 
AM\,1219B does not show any significant asymmetry, just a small overdensity in the north-west arm. There is no significant 
difference between the deconvolved and $S_{2}$ images.

\subsection{Luminosity profile}
\label{profphot}

We derived the $g'$ and $r'$ surface brightness profiles of the $S_{2}$ images to estimate the bulge 
and disc photometric parameters of the pair components. 
The {\sc ISOPHOTE} task of {\sc IRAF/STSDAS} was used \citep{jedrzejewski87}. In this task, the isophotal contour of a
galaxy is fitted with a mean ellipse and parametrized using values
of PA, ellipticity and coordinates of the centre.
The centres of the ellipses were kept fixed while the other parameters were allowed
to vary at each iteration step. During the fitting process,
we adopted a clipping factor of 20\% for the brightest pixels
in each annulus to avoid the pixels of star formation regions. The adopted 
bulge and disc luminosity profiles are the \citet{sersic68} and \citet{freeman70} profiles, respectively :

\begin{equation}
I(r)=I_{b}\exp \left[k_{n} \left( \frac{r}{r_{e}} \right)^{\frac{1}{n}}\right],\\
k_{n}=2n-0.324,
\label{sersiclaw}
\end{equation}

\begin{equation}
I(r)=I_{d} \exp \left[-\left( \frac{r}{r_{d}} \right)\right].
\label{exponentialdisk}
\end{equation}

In \citeauthor{sersic68}'s equation, $I_{b}$ is the central intensity and $r_{e}$ is the effective radius, 
which encloses half of the total bulge luminosity, and the $k$ relation can be estimated for $n\geq1$
with an error lower than 0.1\% \citep{ciotti91}.  $I_{d}$ is the disc central
intensity and $r_{d}$ the scale length of the disc component in \citeauthor{freeman70}'s profile. 

In general terms, we followed the method described by
\citet{aguirre99}, \citet{prieto01} and \citet{cabrera04}. We made a first approximation
assuming that the surface brightness profile  is the sum of bulge and disc
components. Then, the routine begins by fitting the
parameters for the bulge and disc over different ranges of the
profile radius. We fitted the disc
with the \textit{Orthogonal Distance Regression} method \citep{boggs90} using a routine 
in {\sc PYTHON}, and then we subtracted the disc
from the original profile and fitted the bulge component to the
residuals. When the bulge was fitted, we again subtracted
the bulge from the original profile, and the process was
repeated. After some iterations, we obtain the disc and bulge
parameters.

\begin{figure*}
\subfigure{\includegraphics*[width=\columnwidth]{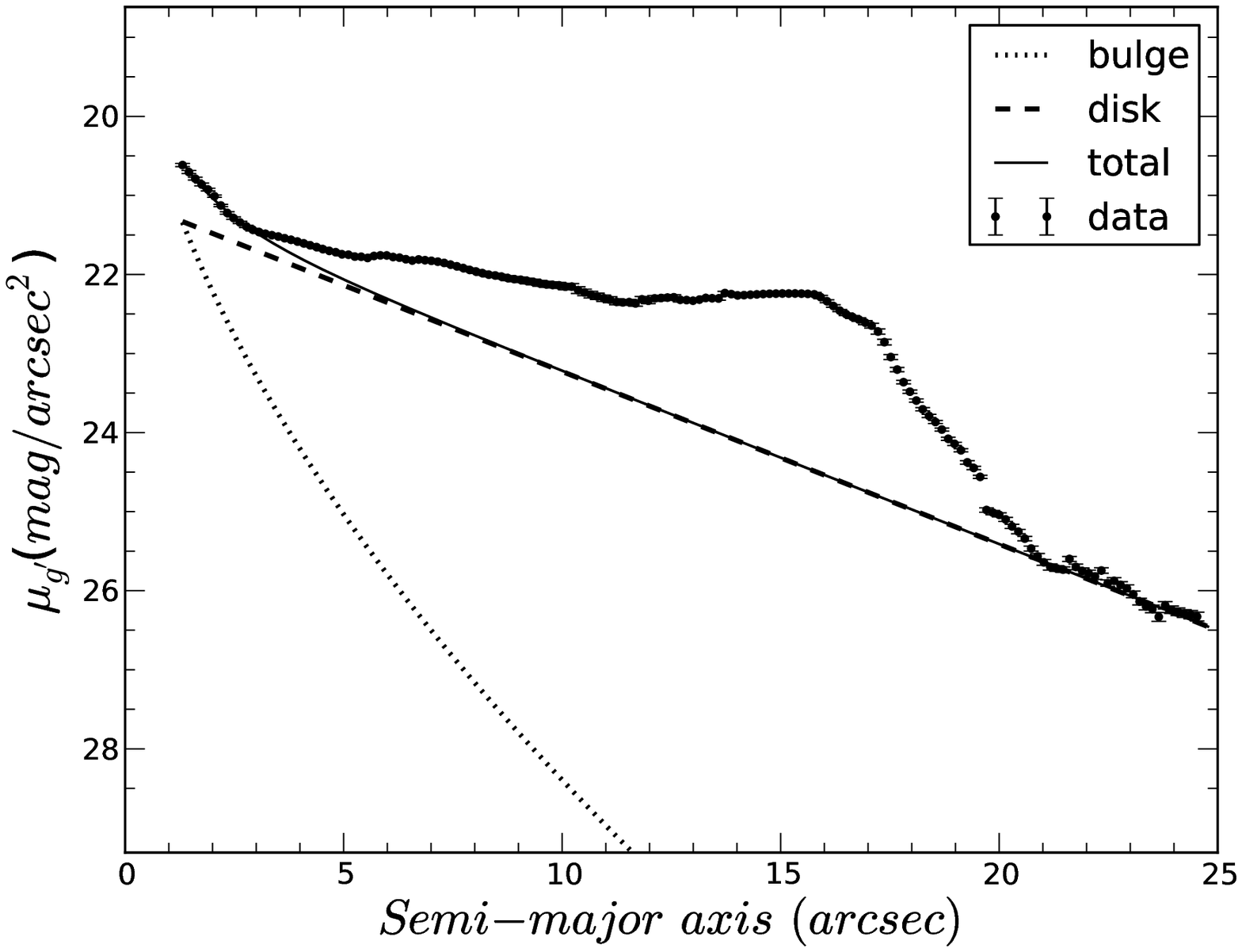}}					     
\subfigure{\includegraphics*[width=\columnwidth]{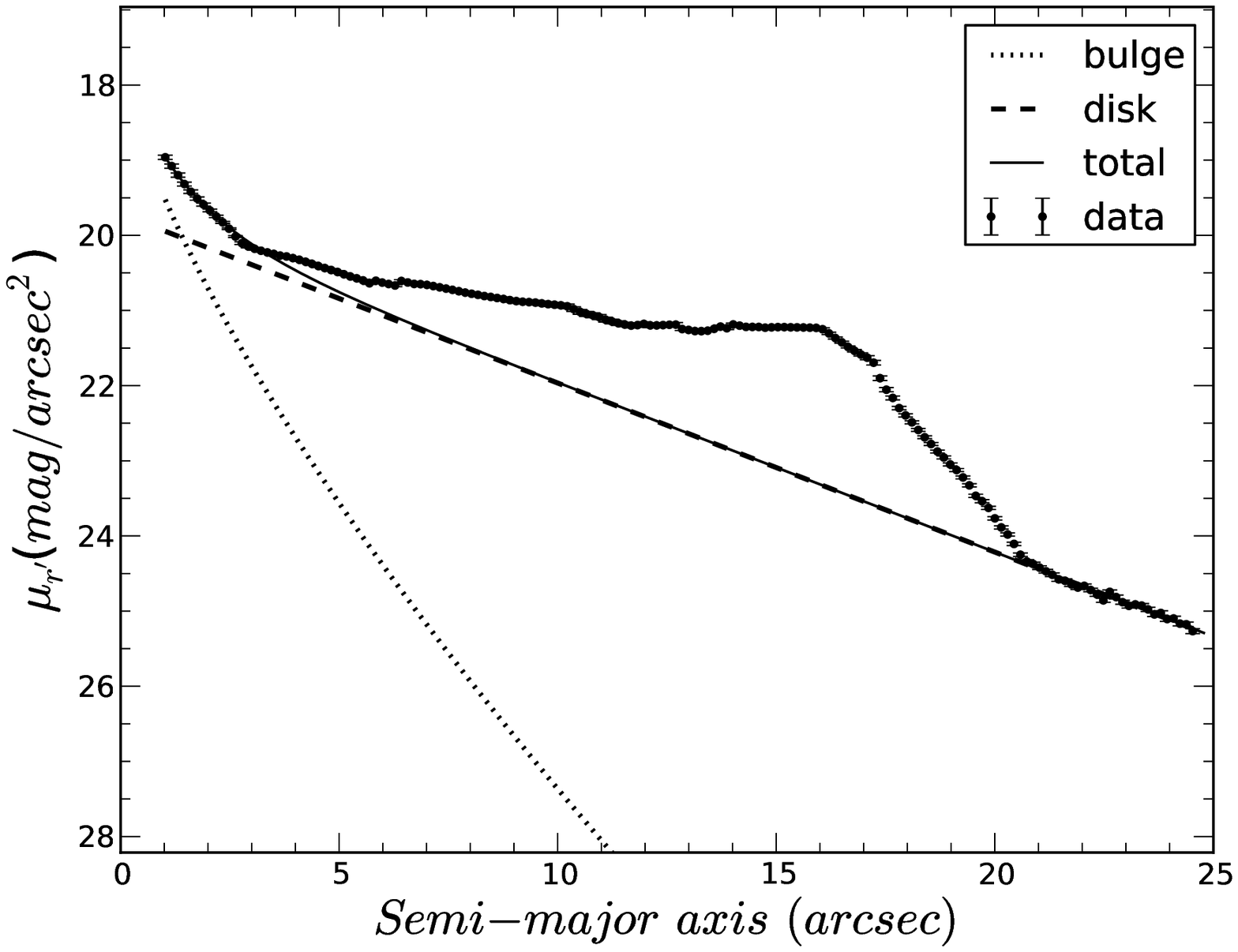}}					     
\caption{Structural decomposition of the surface brightness profiles of AM\,1219A, with $g'$ (left) and $r'$ (right).}
\label{profA}
\end{figure*}

In Fig. \ref{profA}, we present the decomposition of the surface brightness profiles of
AM\,1219A into  bulge and disc structural components in both filters. The sum of these components does 
not fit the observed profile between 2.9 arcsec ($\sim$ 1.3 kpc) and 20.7 arcsec ($\sim$ 9.5 kpc),
where there is an excess of light ($\sim 53 \%$) due to the 
contribution of the star formation along the spiral arm (see Fig \ref{elmeA}). This  excess of $\sim 53 \%$ is 
a typical value for starburst galaxies, and is in agreement with the classification by \citet{pastoriza99} for this galaxy.

The surface brightness profile of the secondary galaxy is decomposed in bulge, disc, and lens-like structures. The latter component was modelled by a \citet{duval83} profile, given by 

\begin{equation}
I(r)=I_{l}\left[1-\left(\frac{r}{r_{l}}\right)^{2}\right].
\label{lens}
\end{equation}

\begin{figure*}
\subfigure{\includegraphics*[width=\columnwidth]{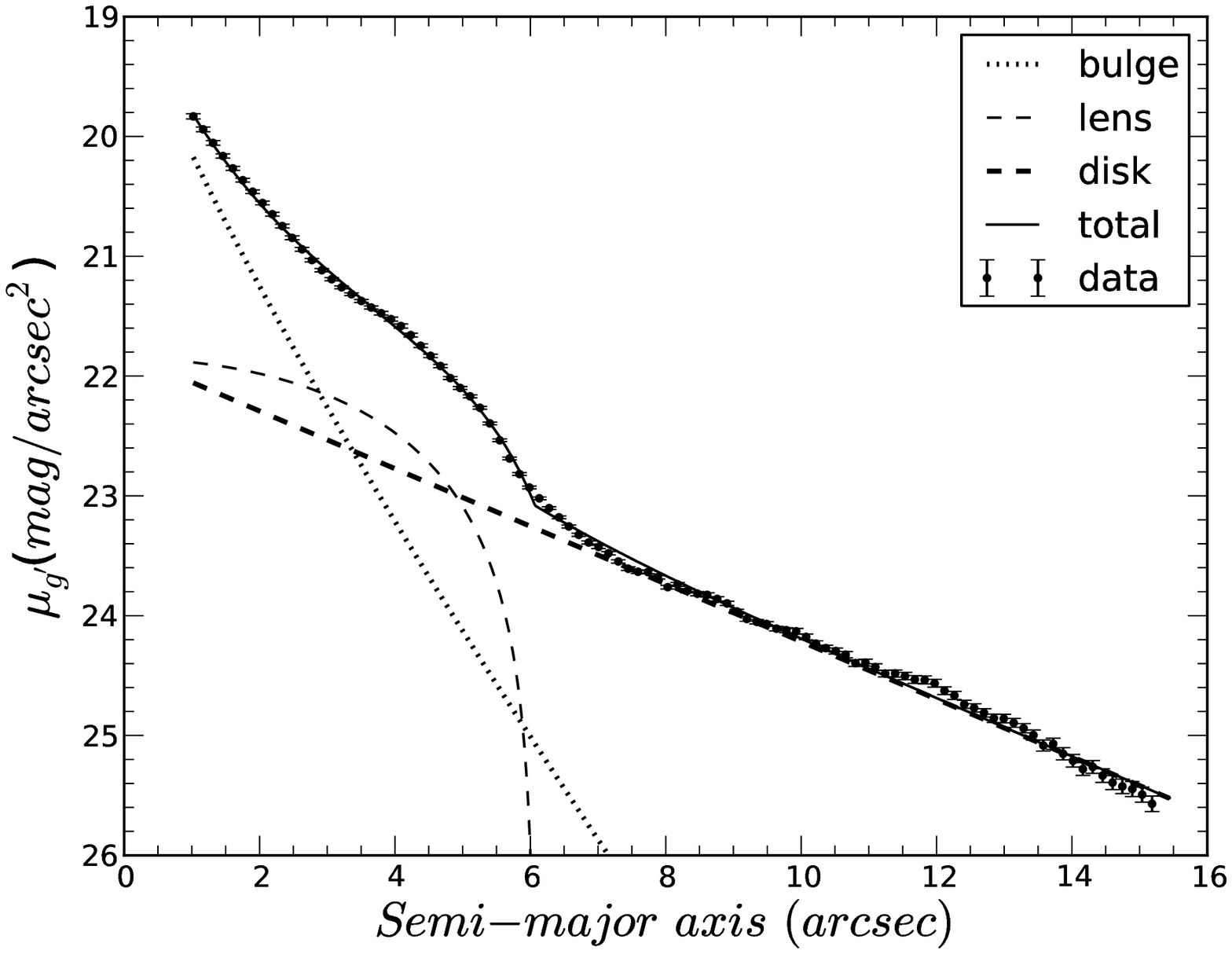}}					     
\subfigure{\includegraphics*[width=\columnwidth]{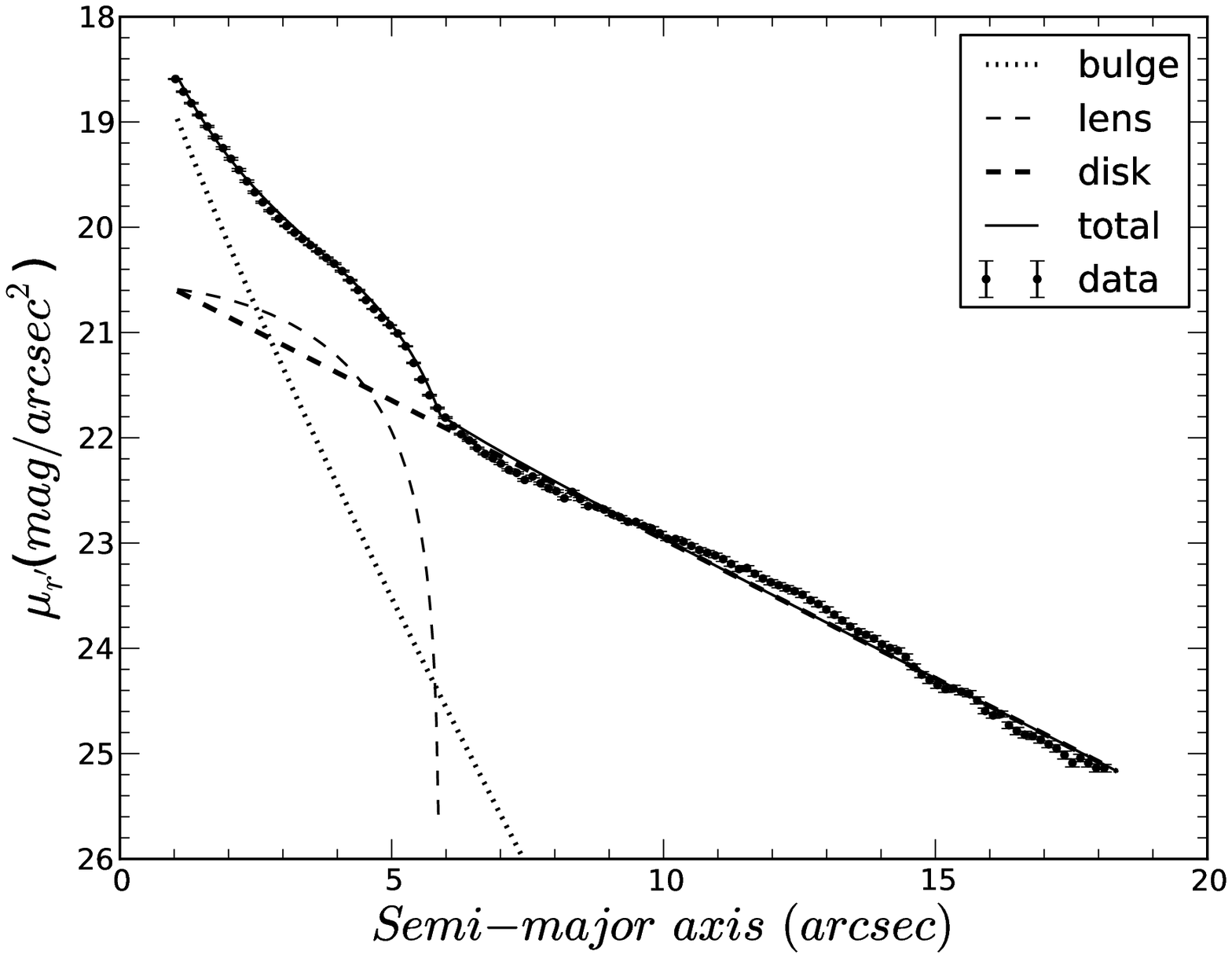}}					     
\caption{Same as Fig. \ref{profA} for AM\,1219B.}
\label{profB}
\end{figure*}

The secondary galaxy surface brightness profile  decomposition is presented in Fig. \ref{profB}. The sum of the adopted components (bulge, disc and lens) fits well the observed profile, with a lower $\chi^2$ of $8.2$ and $8.0$ in the $g'$ and $r'$ bands, respectively.

The structural parameters of AM\,1219A and AM\,1219B in $g'$ and $r'$ are listed 
in Table \ref{photpar}. The scale lengths in the $r'$ band are $2.22\pm0.05$ and 
$1.89\pm0.02$\,kpc for AM\,1219A and AM\,1219B, respectively. Our scale lengths 
and disc central magnitudes agree with the average values: 
$r_{d}=3.8\pm2.1$\,kpc and $\mu_{d}=20.2\pm0.7\,mag/arcsec^{2}$ derived by  
\citet{fathi10a} and \citet{fathi10b} for $\sim30000$  
galaxies with no sign of ongoing interaction or disturbed morphology.  This 
indicates that the interaction has not affected the stellar disc structure. On 
the other hand, the \citeauthor{sersic68} index ($n$) in the bulge structures is 
$<2$ for both galaxies. These values are typical of pseudo bulge 
\citep{kormendy04}. The $r_{e}$ values in $r'$  are $0.76\pm0.02$ and 
$0.68\pm0.05$\,kpc for the main and secondary galaxies, respectively. These 
values agree with the average value for pseudo-bulge found by \citet{gadotti09}. 
The pseudo-bulges are systematically flatter than classical bulges, and their 
morphologies often have nuclear spirals, rings, bars, and patchiness. In addition, 
they have intense star formation activity 
\citep[and reference therein]{kormendy04,fisher10}. These features are in 
agreement with the results in \citet{pastoriza99}, where they found a strong 
star formation in the nuclei of both galaxies; furthermore, AM\,1219A has 
several patches of \ion{H}{ii} regions, as can be seen in the deconvolved image 
in the left-hand panel of Fig. \ref{elmeA}. 

\begin{table*}
\caption{Structural parameters}
\label{photpar}
\begin{tabular}{lcccccccccc}
\noalign{\smallskip}
\hline
\noalign{\smallskip}
       &        & \multicolumn{3}{c}{Bulge} & & \multicolumn{2}{c}{disc} & & \multicolumn{2}{c}{Lens}\\ \cline{3-5}  \cline{7-8} \cline{10-11}
\noalign{\smallskip}
Galaxy & Filter &  $\mu_{b}$ ($mag/$arcsec$^{2}$) & $r_{e}$ (arcsec)& $n$ &  &$\mu_{d}$ ($mag/$arcsec$^{2}$) & $r_{d}$ (arcsec) && $\mu_{l}$ ($mag/$arcsec$^{2}$) & $r_{l}$ (arcsec)\\
 &  [1] & [2] & [3] & [4] & & [5]& [6] & & [7] & [8] \\
\noalign{\smallskip}
\hline
\noalign{\smallskip}
AM\,1219A  & $g^{\prime}$ & $18.11\pm0.91$ & $1.47\pm0.25$ &$1.74\pm0.17$& & $21.04\pm2.21$ & $4.97\pm0.92$ & & - & - \\
         & $r^{\prime}$ & $17.60\pm0.70$ & $1.64\pm0.04$ &$1.41\pm0.16$& & $19.72\pm0.27$ & $4.82\pm0.11$   & & - & - \\
AM\,1219B  & $g^{\prime}$ & $18.74\pm1.81$ & $1.76\pm0.34$ &$1.20\pm0.47$& & $21.81\pm0.07$ & $4.51\pm0.05$ & & $21.01\pm0.07$ & $6.07\pm0.16$\\
         & $r^{\prime}$ & $17.33\pm1.57$ & $1.48\pm0.12$ &$1.17\pm0.34$& & $20.33\pm0.08$ & $4.11\pm0.04$ & & $20.56\pm0.33$ & $5.89\pm0.06$ \\
\noalign{\smallskip}
\hline
\noalign{\smallskip}
\end{tabular}
\end{table*}

In addition to the analysis of the surface brightness profiles, we determined the position angle (PA) and inclination ($i$) of the discs of both galaxies, assuming the mean PA and $i$ values of the most external ellipses, where the disc component dominates the radial profile and the isophotes are not affected by the spiral structure. PA and $i$ calculated are listed in Table \ref{ipapar}.

\begin{table}
\caption{Inclination and position angle }
\label{ipapar}
\begin{tabular}{lcc}
\noalign{\smallskip}
\hline
\noalign{\smallskip}
Galaxy & $i$ (\degr)  & PA  (\degr)  \\
\noalign{\smallskip}
\hline
\noalign{\smallskip}
   AM\,1219A    & $39.8\degr\pm1.5\degr$    & $29.0\degr\pm0.6\degr$ \\
   AM\,1219B    & $49.3\degr\pm0.2\degr$    & $-42.0\degr\pm0.7\degr$  \\

\noalign{\smallskip}
\hline
\noalign{\smallskip}
\end{tabular}
\end{table}

\section{Ionized gas kinematics}
\label{vel}

Radial velocities were derived by measuring the centroids of Gaussian curves fitted to the profiles of the strongest
emission lines ([\ion{N}{ii}] $\lambda 6548.04$, $\rm H\alpha$ $\lambda 6563$, [\ion{N}{ii}] $\lambda 6584$ 
and [\ion{S}{ii}] $\lambda 6717$). For spectra with  very low signal-to-noise ratio,  only  the $\rm H\alpha$ 
line was used to determine the radial velocity. The final radial velocity for each spectrum was obtained by 
averaging the individual measurements from the detected emission lines, and the errors correspond to the 
standard deviation of these measurements around the mean. For radial velocities that were measured with a 
single emission line, we used \citeauthor{keel96}'s expression \citep{keel96} to determine  
the uncertainty. Then, we subtracted from them the systemic velocity. We adopted the radial velocity of 
the central aperture of each galaxy as systemic velocity, which are 6932 and 6985  km s$^{-1}$ for 
the main and secondary galaxies, respectively. \citet{donzelli97} estimated systemic velocities of 
6957 and 6979 km\,s$^{-1}$ for AM1219A and AM1219B, respectively. Our estimates agree to within 
1 \% with those previous determinations. 

\begin{figure*}
\subfigure{\includegraphics*[width=\columnwidth]{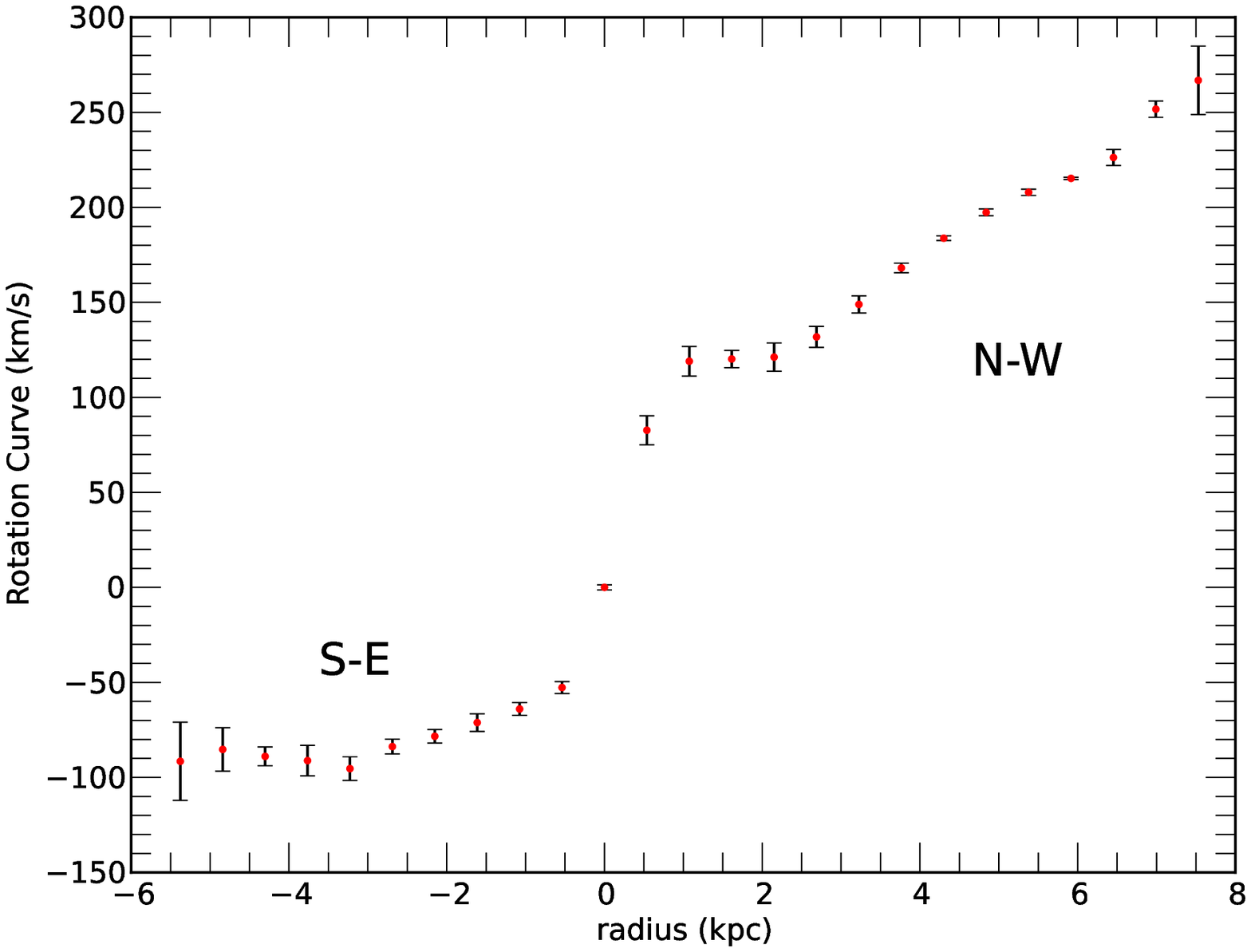}}					     
\subfigure{\includegraphics*[width=\columnwidth]{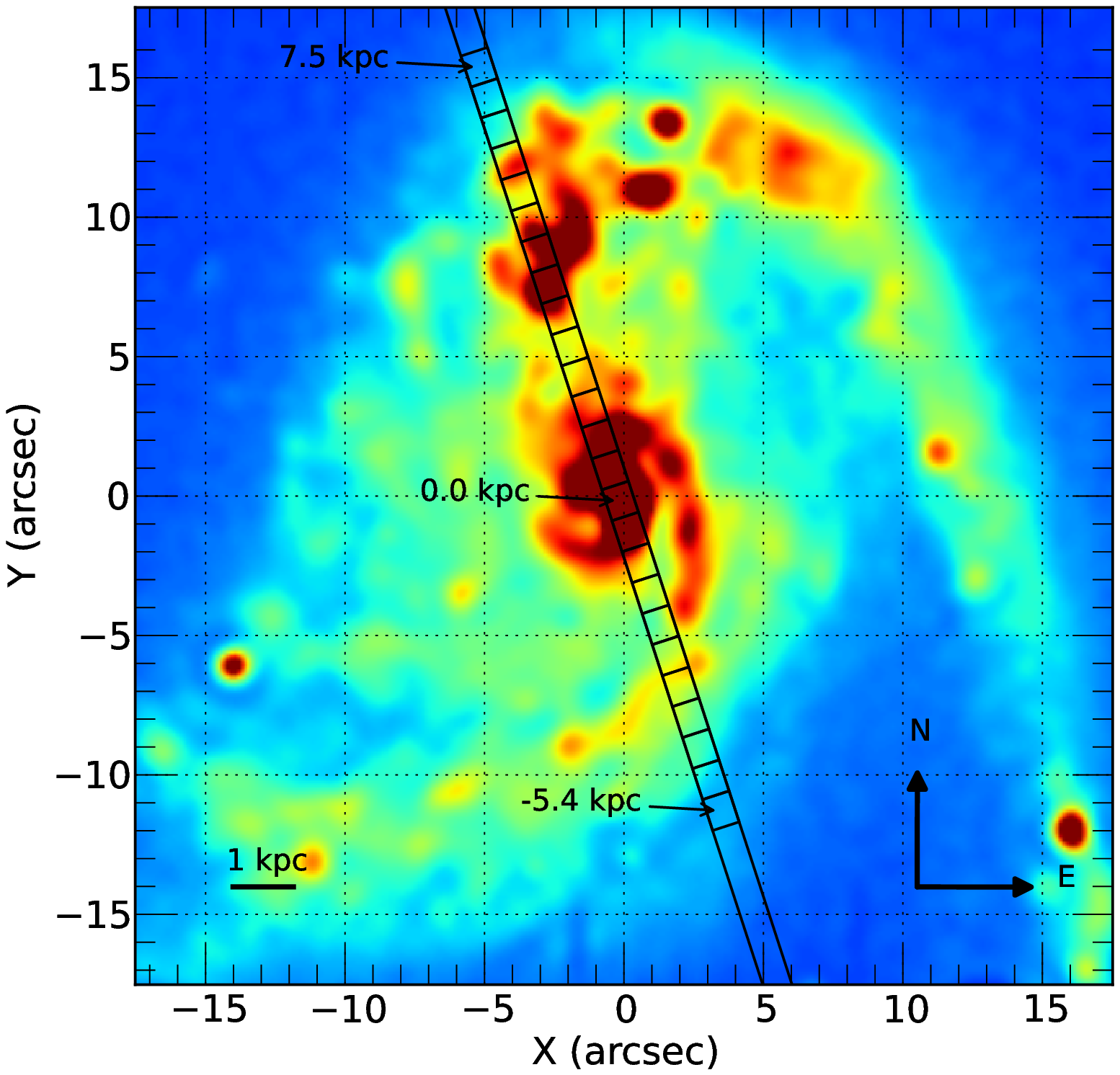}}					     
\caption{Left: kinematics along PA=162\degr in AM\,1219A. The velocity scale corresponds to the observed values after subtraction of the systemic velocity for each galaxy, without correction for inclination on the plane of the sky. Right: the apertures extracted along the slit.}
\label{curveA}
\end{figure*}

Fig. \ref{curveA} shows the observed rotation curve of AM\,1219A  and the $r'$ 
image with the long-slit  apertures. The rotation curve is quite asymmetric, 
with the south-east (S-E) side being typical of discs, increasing from the 
centre until reaching a peak at $\sim$ 3.2 kpc and flattening at 
$\sim -84$ km s$^{-1}$. On the other hand, the north-west (N-E) side has the 
same  behaviour as the S-E up to $\sim$ 3.5 kpc, then increasing to reach the 
velocity of $\sim$ 250 km s$^{-1}$ at 8.2 kpc. The highest velocities are 
spatially coincident  with the N-E \ion{H}{ii} region complexes 
(see Fig. \ref{curveA}). The non-symmetric behaviour in the rotation curves has 
also been observed in other interacting pairs  
\citep[e.g.,][]{rubin91,rubin99,dale01,mendes03,fuentes04,presotto10} and 
predicted by numerical simulation of \citet{pedrosa08} and \citet{kronberger06} 
for major and minor mergers, respectively.

\begin{figure*}
\subfigure{\includegraphics*[width=\columnwidth]{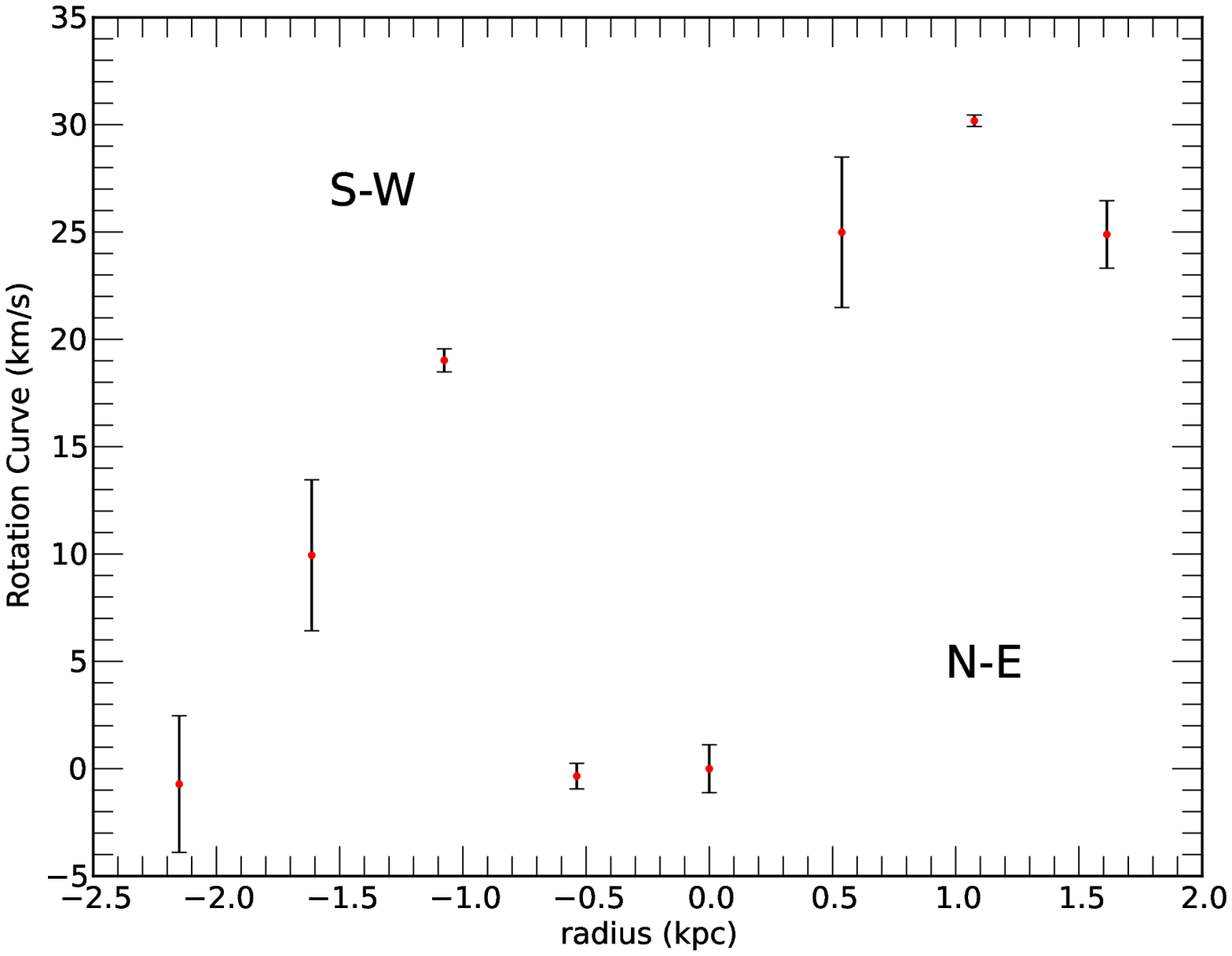}}					     
\subfigure{\includegraphics*[width=\columnwidth]{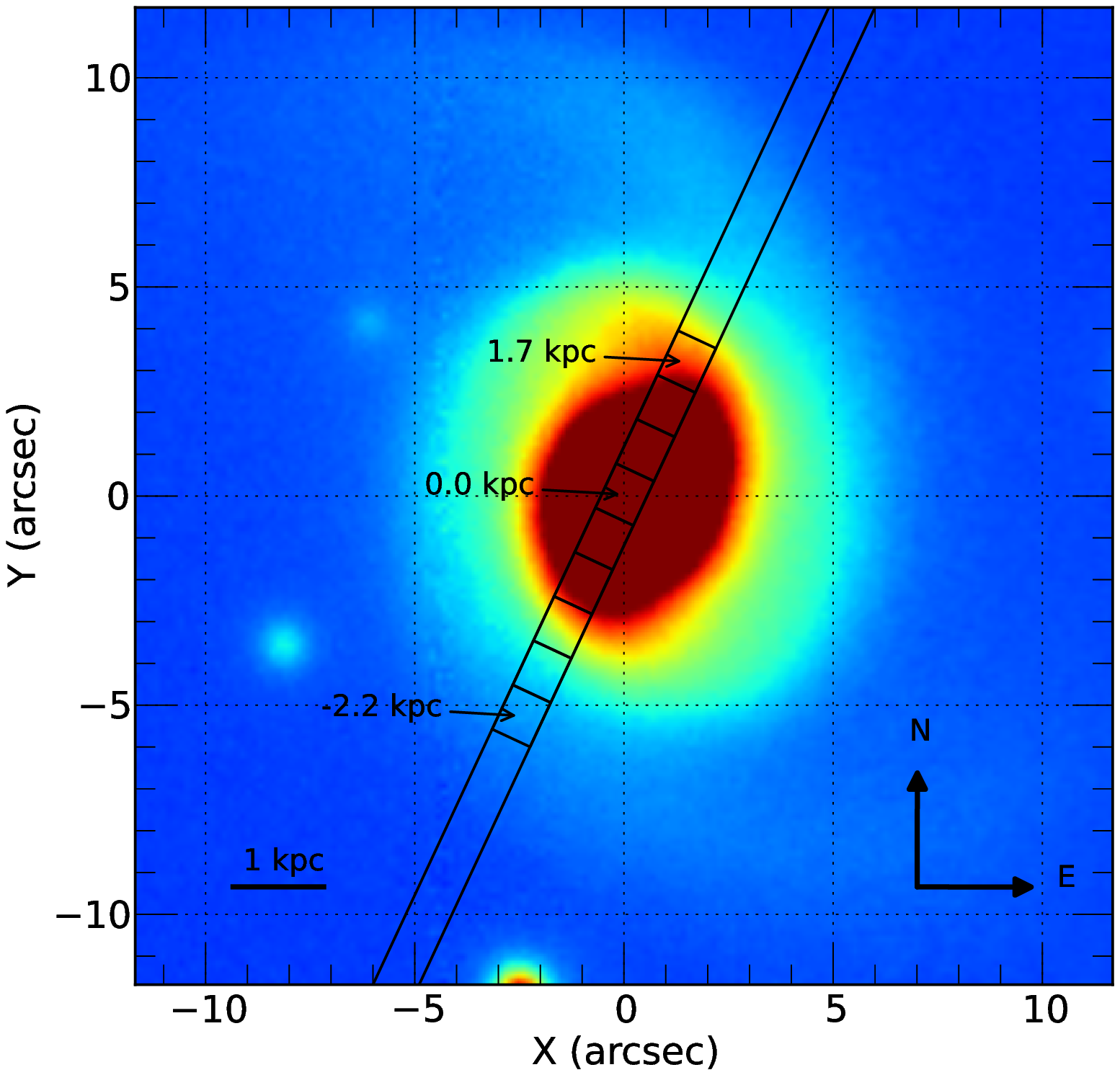}}					     
\caption{Same as Fig. \ref{curveA}, for AM\,1219B and a slit with a PA=$25\degr$}
\label{curveB}
\end{figure*}

The observed rotation curve  and the $r'$ image with the long-slit apertures for AM\,1219B are shown in Fig. \ref{curveB}. Clearly,
AM\,1219B does not present a well-defined and symmetric rotation curve along the observed slit position. 
The decomposition of the surface brightness profile (Section \ref{profphot})  shows that AM\,1219B  
is a disc galaxy. Thus, the fact that we do not detect rotation may be due to the long-slit 
position (PA =$25\degr$) lying almost perpendicular to the line of nodes ($\sim-42\degr$, see Table \ref{ipapar}).

\section{Rotation curve models}
\label{massmodel}

The mass distribution in AM\,1219A is modelled as sum of the bulge, disc and dark halo components. The bulge 
and disc mass distributions are assumed to be given by their respective luminosity distributions to within 
an unknown M/L \citep{kent87}. If we assume that the mass distribution of AM\,1219A follows the deprojected 
luminosity distribution with constant M/L for each component, then, we can obtain the mass surface density 
for bulge and disc from the \citeauthor{sersic68} and \citeauthor{freeman70} surface brightness profiles, 
respectively. The halo mass distribution can be either parametrized or derived from the observed rotation 
curve. Below we will describe in some detail the mass model for the components and  give the expressions for 
the circular velocity ($V_c$).

For the bulge mass distribution, we use the rotation curve derived for a \citeauthor{sersic68} profile density. This profile is obtained  by an Abel integral equation \citep{binney87,simonneau04}, which relates surface brightness (equation \ref{sersiclaw}) to density:

\begin{equation}
\label{density_sersic}
\rho (s) =  \frac{1}{\pi}\frac{k_{n}}{n}I_{b}\Upsilon_{b}\int_{s}^{\infty}\frac{exp[-k_{n}z^{\frac{1}{n}}]z^{\frac{1}{n}-1}}{\sqrt{z^{2}-s^{2}}}\, {\rm d}z,
\end{equation}
where $I_{b}$, $r_{e}$, $n$  and $k_{n}$ are those in equation \ref{sersiclaw}, and $s=(r/r_{e})$. $\Upsilon_{d}$ is the M/L for the bulge component. For $n>1$, the integration of this equation has no analytical solution. Thus, we will use an analytical approximation proposed by \citet{simonneau04} for $\rho (s)$ that allows an easy computation of the mass and gravitational potential to any required degree of precision:  

\begin{equation}
\label{curv_simmoneau}
\rho (s) =  \frac{k}{\pi}\frac{2}{n-1}\frac{1}{s^{\frac{n-1}{n}}}I_{b}\Upsilon_{b}\int_{0}^{1}\frac{exp[-ks^{\frac{1}{n}} (1-x^2)^{-\frac{1}{n-1}}]}{\sqrt{1-(1-x^2)^{-\frac{2n}{n-1}}}}x \, {\rm d}x.
\end{equation}

Equation \ref{curv_simmoneau} can be solved by Gaussian numerical integration. The cumulative mass profile is given by

\begin{equation}
\label{masssersic}
M(r)=4\pi\int_0^r r^2\rho(r)\, {\rm d}r. 
\end{equation}

For a spherical density distribution with \citeauthor{sersic68} profile, we may readily evaluate the circular velocity ($V_{b}$) by equating the gravitational attraction  to the centripetal acceleration:

\begin{equation}
\label{curv_sersic}
V_{b}^{2}(r)=G\frac{M(r)}{r}.
\end{equation}

For the disc, the cumulative mass of the exponential profile \citep{binney87} is 

\begin{equation}
\label{mass_freeman}
M(r)=2\pi\Upsilon_{d}I_{d}r_{d}^{2}[1-\exp(-r/r_{d})(1+r/r_{d})],
\end{equation}
where $I_{d}$ and $r_{d}$ are those in equation \ref{exponentialdisk} and $\Upsilon_{d}$ is the M/L for disc component. The circular velocity ($V_{d}$) curve derived for an exponential disc is given by the following  equation \citep{freeman70}:

\begin{equation}
\label{curv_freeman}
V_{d}^{2}(r)=4\pi G \Upsilon_{d}I_{d}r_{d}y^{2}[I_{0}(y)K_{0}(y)-I_{1}(y)K_{1}(y)],
\end{equation}
where $y=r/2r_{d}$, $I_{n}$ and $K_{n}$ are modified Bessel functions of the first and second kinds, respectively. $G$ is the gravitational constant.

For the halo mass model, we use the density profile proposed by \citeauthor{navarro95}  \citeyearpar[ hereafter NFW]{navarro95,navarro96,navarro97}. In this case the dark matter density is given by  

\begin{equation}
\label{profnfw}
\rho (r) =\frac{\rho_{0}\rho_{c}}{(\frac{r}{r_{s}})(1+\frac{r}{r_{s}})},
\end{equation}
where $r_s$ is a characteristic radius, $\rho_{0}$ is the characteristic overdensity and 
$\rho_{c}=3H_0^{2}/8 \pi G$  is the critical density ($H_0$ is the current value of Hubble's constant).
The virial halo mass is usually in the literature as $M_{200}=\frac{4}{3}\pi200\rho_{c}r^3_{200}$.
It is useful to define a dimensionless parameter $c\equiv r_{200}/r_s$, where $c$  is called the halo concentration. 
$\rho_{0}$ can be linked to $c$ by \citep{navarro96}

\begin{equation}
\label{pho_c}
\rho_0 =\frac{200}{3}g(c);\,\, g(c)=\frac{c^3}{[\ln(1+c-c/(1+c))]}.
\end{equation}

The circular velocity ($V_h$) in the NFW profile is

\begin{equation}
\label{curv_nfw}
V^2_h(r)=\frac{GM_{200}}{g(c)r}\left[\ln(1+cr/r_{200})-\frac{cr/r_{200}}{1+cr/r_{200}}\right].
\end{equation}

Then, we can calculate $V_h$ for the NFW profile via the concentration ($c$) and 
virial mass ($M_{200}$).

The final rotation curve model is computed from the squared sum of the circular velocities of the bulge, disc and halo components:

\begin{equation}
\label{curv_final}
V_{c}^{2}(r)=V_{b}^{2}(r)+V_{d}^{2}(r)+V_{h}^{2}(r).
\end{equation}

This equation has four free parameters: the bulge and disc M/L ratios, $\Upsilon_{b}$ and $\Upsilon_{d}$, respectively, and the halo parameters,  $M_{200}$ and $c$. The other parameters are photometric, which were determined in Section \ref{profphot} for the bulge ($I_{b}$, $r_{e}$ and $n$) and  the disc ($I_{d}$ and $r_{d}$) components. We use photometric parameters of the $r'$ band, because it provides a better estimate of the underlying mass distribution than $g'$.

\section{Mass models}
\label{fitmodel}

\begin{figure*}
\subfigure{\includegraphics*[width=\columnwidth]{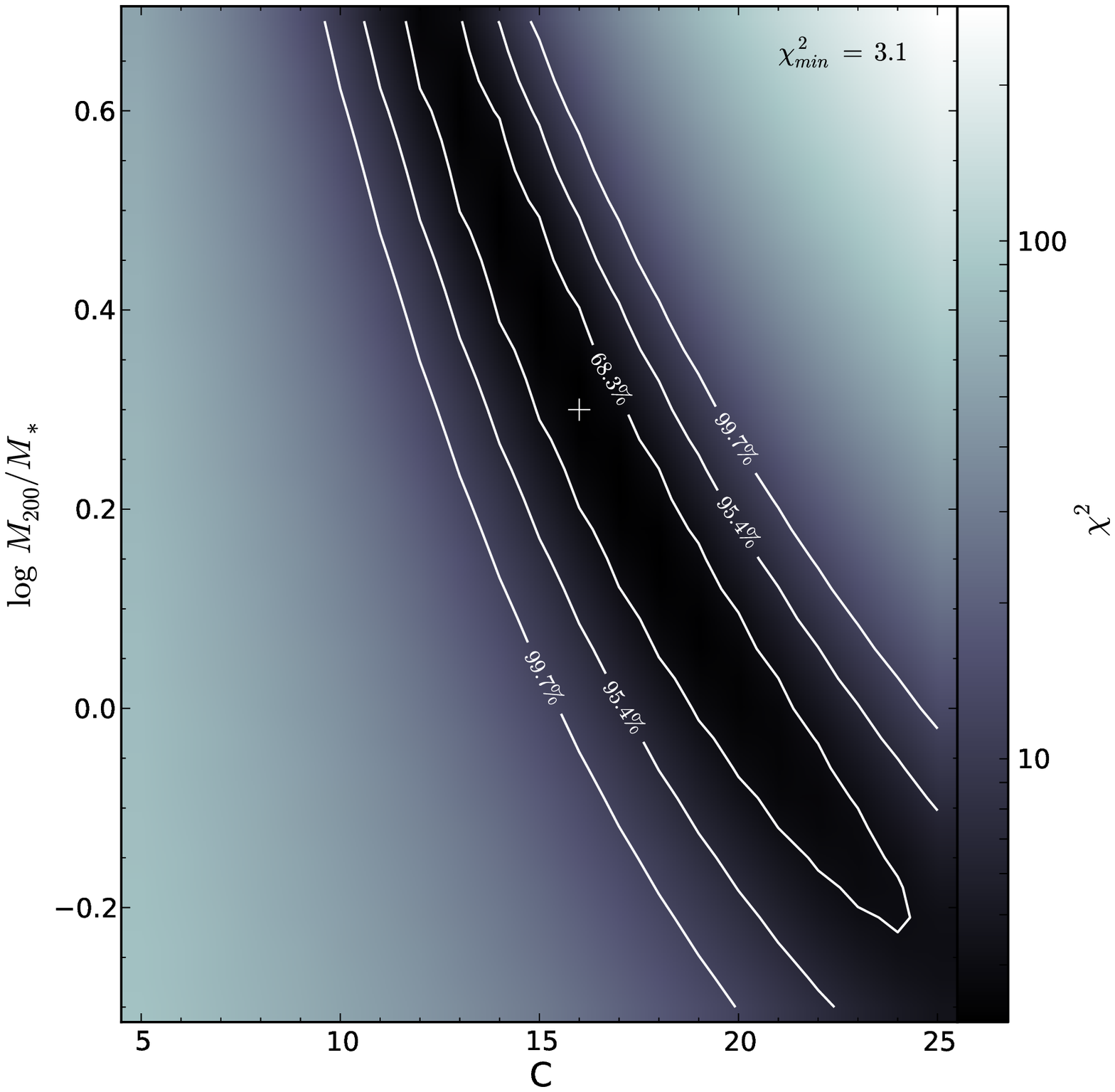}}					     
\subfigure{\includegraphics*[width=\columnwidth]{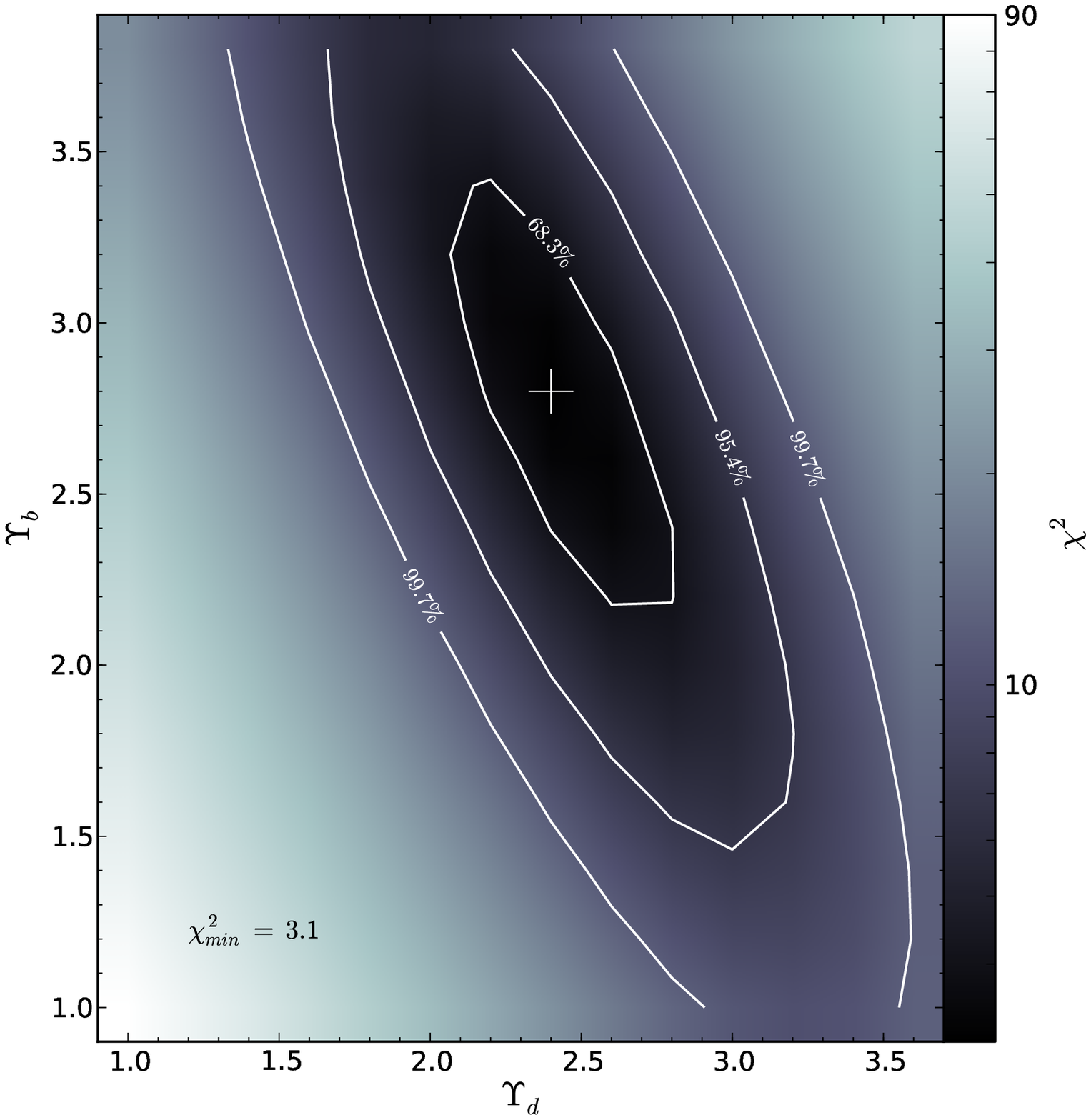}}					     
\caption{Left: the $\chi^2$ space projections in the plane 
$\log(M_{200}/M_{\ast})$--$c$ by fixing on the global minimum values for M/L ratios 
($\Upsilon_{b}=2.8$,  $\Upsilon_{d}=2.4$). Right: the $\chi^2$ space projections 
in the plane $\Upsilon_{b}$--$\Upsilon_{d}$ by fixing on the global minimum 
values for the NFW halo parameters ($\log(M_{200}/M_{\ast})=0.3$, $c=16.0$). 
Contours of $\Delta \chi^2$ corresponding to a probability of 68.3, 95.4 and  
99.7 per cent (1$\sigma$, 2$\sigma$, 3$\sigma$) for 1 degree of freedom. The 
plus symbol indicates the global minimum $\chi^2$.}
\label{chimap}
\end{figure*}

The literature presents basically  two approaches to study the mass distribution: 
the maximum-disc \citep[and references therein]{vanalbada85,carignan85} and the 
best-fitting method \citep[and references therein]{kent87}. The first approach 
assumes that the stellar component (bulge+disc) dominates the mass distribution 
in the inner part. The bulge and disc  M/L ratios are constrained by the 
rising part of the rotation curve. Any excess in rotation velocity at large radii 
is then attributed to a dark halo component. On the other hand, the second 
approach uses a $\chi^2$ analysis to determine the combination of parameters 
that best fits the observed rotation curve.

In this paper, we adopt a general method that includes all possible distributions 
of values between stellar and dark matter masses. We explore a range of values of 
$M_{200}$ and $c$  halo  parameters for a grid  of M/L ratios of the bulge 
($1.0\leq\Upsilon_{b}\leq 3.6$) and disc ($1.0\leq\Upsilon_{d}\leq 3.6$). 
We adopt  the following ranges  for the halo parameters: 
$-0.3\leq \log (M_{200}/M_{\ast})\leq0.69$ 
(where $M_{\ast}=1.0\times10^{12}\,\mbox{M}_{\odot}$) and $5.0\leq c \leq 25.0$ 
\citep{navarro96,bullock01}. Thus, for these parameters, the  minimum disc 
corresponds  to $\Upsilon_{b}=1.0$ and  $\Upsilon_{d}=1.0$; the  maximum disc, 
however,  corresponds to $\Upsilon_{b}=3.6$ and  $\Upsilon_{d}=3.6$. 
We developed a {\sc PYTHON} code to compute for each  point of the $\chi^2$ space 
($\Upsilon_{b}$, $\Upsilon_{d}$, $\log(M_{200}/M_{\ast})$ and $c$), the circular 
velocity for each component, the resulting  $V_{c}$  given by 
equation \ref{curv_final} and the $\chi^2$ model of the rotation curve.

We adopted as observed velocity curve the points along the S-E side of AM\,1219A because, as discussed 
in Section \ref{vel}, the N-W side of AM\,1219A is significantly affected by the presence of AM\,1219B. 
In contrast, the S-E side of the galaxy seems to be relatively undisturbed. The adopted velocity curve 
has been corrected for Galactic systemic motion, inclination with respect to the plane of the sky 
($i=40\degr$), and PA of the line of nodes (PA$=29\degr$) (see Table \ref{ipapar}).

The best-fitting model corresponding to the global minimum of $\chi^2$ space  was found for the
following parameters: $\Upsilon_{b}=2.8$, $\Upsilon_{d}=2.4$, $\log(M_{200}/M_{\ast})=0.3$ and $c=16.0$. In order to 
illustrate the global minimum and its convergence pattern, we present in Fig. \ref{chimap} the $\chi^2$ 
space projections in the planes $\log(M_{200}/M_{\ast})$--$c$ and $\Upsilon_{b}$--$\Upsilon_{d}$ by fixing on the 
global minimum values. The convergence patterns are smooth and the global minimum is easily identifiable 
in both projections. The contours indicate the degree of the correlation between parameters. 
Then, as can be seen in the left-hand panel in Fig. \ref{chimap}, there is a strong correlation between $M_{200}$ and $c$, 
which is because there is a degeneracy between them: a decrease in  $c$ is balanced with an increase in 
$M_{200}$, and vice versa. This is due to the rotation curve that does not extend to large radii.

\begin{figure}
\includegraphics[width=\columnwidth]{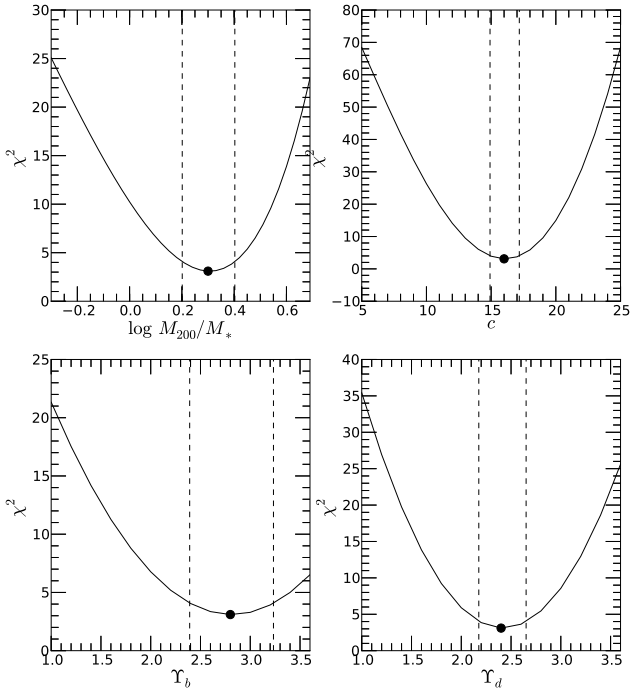}					     
\caption{One-dimensional projection of $\chi^2$ space in the fit parameters: 
$\log(M_{200}/M_{\ast})$, $c$, $\Upsilon_{b}$, and $\Upsilon_{d}$. 
The absolute minimum is shown by the black circle, and the vertical dashed lines 
show the 1$\sigma$ confidence bands quoted for the parameter uncertainties.}
\label{chi1d}
\end{figure}

Fig. \ref{chi1d} shows the one-dimensional  projections of the $\chi^2$ space in the fit parameters: 
$\log(M_{200}/M_{\ast})$, $c$, $\Upsilon_{b}$, and $\Upsilon_{d}$. Each plot is produced by taking global 
minimum values for all the parameters except  for the parameter chosen for the projection, which are 
allowed to vary within the ranges specified above. Then, we compute $\chi^2$  for the adopted range, 
but keeping the remaining parameters fixed. In these plots, the $\Delta\chi^2$ variation obeys the 
$\chi^2$ probability distribution for 1 degree of freedom. So $\Delta\chi^2=1$ corresponds to 1$\sigma$ 
uncertainty or 68\%. In Table \ref{parmodels} are listed the parameters and their errors (at the $1\sigma$ level) 
for the model with the global minimum of $\chi^2$, and the cumulative mass for each component and 
their total sum inside 6.2 kpc, which is the radius of the last velocity point observed.

\begin{table*}
\caption{Mass model parameters for AM\,1219A}
\label{parmodels}
\begin{tabular}{cccccccc}
\noalign{\smallskip}
\hline
\noalign{\smallskip}
 $\Upsilon_{b}$ & $\Upsilon_{d}$ & $c$ &  $M_{200}/\mbox{M}_{\odot}$ &  $M_b/\mbox{M}_{\odot}$ & $M_d/\mbox{M}_{\odot}$ & $M_h/\mbox{M}_{\odot}$ & $M_t/\mbox{M}_{\odot}$\\
\hline
\noalign{\smallskip}
$2.8_{-0.4}^{+0.4}$ & $2.4_{-0.2}^{+0.3}$ & $16.0_{-1.1}^{+1.2}$ &$2.0_{-0.4}^{+0.5}\times10^{12}$ & $5.8\times10^{9}$ & $2.5\times10^{10}$ & $5.2\times10^{10}$ & $8.3\times10^{10}$ \\   
\noalign{\smallskip} 
\hline    
\noalign{\smallskip}
\end{tabular}
\end{table*}

\begin{figure}
\centering
\includegraphics[width=\columnwidth]{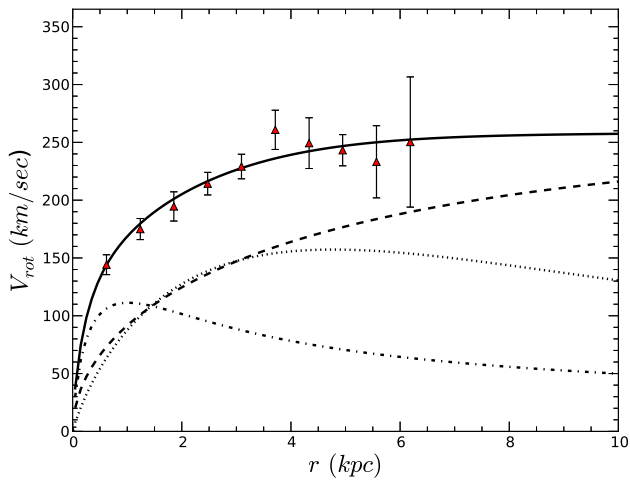}  				     
\caption{The observed rotation curve (red triangles with error bars) of AM\,1219A corrected by the galaxy inclination 
overlaid on the best-fitting model of rotation curve  (continuous line) and their components: 
bulge (dash-dotted), disc (dotted), and halo (dashed).}
\label{modelbest}
\end{figure}

The observed rotation curve overlaid on the best-fitting model is shown in Fig. \ref{modelbest}. The model provides a good 
match to the observed curve over all radii,  with $\chi^2$ of $3.1$. The bulge component dominates the inner part 
($r$ $\lesssim$ $1.5$ kpc) of the rotation curve. The disc and halo have equal weights in the 
middle part ($1.5\lesssim r \lesssim 3.5$ kpc), and the halo component  becomes dominant in the outer 
parts  ($3.5 \lesssim r$ kpc), where the curve is flat.

\begin{table}
\caption{Comparison of AM\,1219A halo parameters with those found in other galaxies}
\label{compmodels}
\begin{tabular}{lccccc}
\noalign{\smallskip}
\hline
\noalign{\smallskip}
Galaxy  & $c$ & $R_{200}$ (kpc) &  $M_h/\mbox{M}_{\odot}$ \\
\hline
\noalign{\smallskip}
AM\,1219A ($\chi^2_{\scriptsize{\mbox{min}}}$)    & 16 & 184  &  $2.0_{-0.4}^{+0.5}\times10^{12}$ \\
Milky Way $^{(a)}$            & 18 & 186  &  $0.8_{-0.2}^{+1.2}\times10^{12}$ \\
M\,31 $^{(b)}$                & 13 & 200  &  $1.04\times10^{12}$ \\
Simulation Sc $^{(c)}$        & 22 & 239  &  $0.79\times10^{12}$ \\
\noalign{\smallskip} 
\hline
\noalign{\smallskip}
\end{tabular}
{\bf Note:} values taken from, $^{(a)}$ \citet{battaglia05}, $^{(b)}$ \citet{tamm12} and 
$^{(c)}$ ERIS simulation for the formation of late type galaxies \citep{guedes11}.

\end{table}

In Table \ref{compmodels}, we compare the NFW halo parameter obtained for AM\,1219A using 
our general method with those found for Milky Way  \citep{battaglia05}, M31 \citep{tamm12}, 
and ERIS simulation for the formation of  late  type galaxies \citep{guedes11}. 
Milky Way and M31 are typical late type galaxies, and we can see that their halo parameters are 
close to those found for  AM\,1219A.

We use the model rotation curve to extrapolate the mass profile at the equivalent radius (10.6\,kpc) 
of the outermost isophote ($\mu_{B}=25.2\,mag/$arcsec$^{2}$, see Table \ref{magpar}), finding a 
mass of $1.6\times10^{11}\, \mbox{M}_{\odot}$. For the same region, we calculate, $L_{B}=1.8\times10^{10}\,\mbox{L}_{\odot}$, 
which gives M/L=8.9. This result is higher compared to the mean value, $M/L_{B}=4.7\pm0.4$, for 
isolated galaxies of the same Hubble type (Sbc) obtained by \citet{faber79}. However, it agrees 
with the mean value, $M/L_{B}=8.5\pm1.1$, derived by \citet{blackman84} for binary galaxies.

\section{Conclusions}
\label{final}
We present an observational study on the  effects of interactions in the morphology and  kinematics of the galaxy pair AM\,1219-430. 
The data consist in $g'$ and $r'$ images  and long-slit spectra in the  wavelength range from 4\,280 to 7\,130\AA\ , obtained with  
the  GMOS at Gemini South. The main findings are the following.
\begin{enumerate}

\item  We detected a striking bridge of material connecting both galaxies, together with a tidal tail in the secondary galaxy.

\item The total apparent and absolute magnitudes in $g'$ and $r'$ for each galaxy were measured. The pair has  $\Delta M_{g'}=1.52$ and $\Delta M_{r'}=1.41$, which means that in luminosity, the primary galaxy is about 3.8 times brighter than the secondary.

\item The symmetric and non-symmetric parts of AM\,1219A and AM\,1219B were 
separated using the symmetrization method of \citet{eem92}. The  symmetric 
images show what should be the ``original disc'' and the non-perturbed spiral 
pattern. From these images, we calculate  the PA and $i$ for both galaxies. We 
found for AM\,1219A  a PA=$29.0\degr\pm0.6\degr$ and an $i=40.8\degr\pm1.8\degr$, 
and  AM\,1219B has a PA=$-42.0\degr\pm0.7\degr$ and an $i=49.3\degr\pm0.1\degr$. 
On the other hand, the non-symmetric image of AM\,1219A shows a one-arm 
structure, with several \ion{H}{ii} region complexes along it. 

\item The surface brightness profile of AM\,1219A was decomposed into bulge and disc components. The profile shows a light excess of $\sim 53 \%$ due to the contribution of star-forming regions in the arms, which is typical of starburst galaxies. On the other hand, the surface brightness profile of AM\,1219B reveals the existence a lens structure in addition to the  bulge and disc. 

\item The scale lengths  and central magnitudes of the disc structure of the galaxies agree with the average values derived for galaxies with no sign of ongoing interaction or disturbed morphology \citep{fathi10a,fathi10b}. This indicates that the interaction has not affected the disc structure.  The \citeauthor{sersic68} index ($n<2$) and the effective and scale radii in the bulge structures of both galaxies are typical of pseudo-bulge \citep{kormendy04,gadotti09}.    

\item The rotation curve of AM\,1219A, derived from the ionized gas emission 
line, is quite asymmetric, suggesting that the gas kinematic was perturbed by 
interaction. The highest velocities of north-west side of the rotation curve 
are spatially coincident  with the \ion{H}{ii} region complexes. 

\item  A general method was used in order to explore all possible values of 
the stellar and dark matter masses. The overall best-fitting solution for the 
mass distribution of AM\,1219A was found with M/L for bulge and disc of 
$\Upsilon_{b}=2.8_{-0.4}^{+0.4}$ and $\Upsilon_{d}=2.4_{-0.2}^{+0.3}$, 
respectively, and  NFW halo parameters of 
$M_{200}=2.0_{-0.4}^{+0.5}\times10^{12}\, \mbox{M}_{\odot}$ and $c=16.0_{-1.1}^{+1.2}$. 
The estimated dynamical mass is $1.6\times10^{11}\, \mbox{M}_{\odot}$ within a radius 
of $\sim10.6$ kpc. M/L$_B$ at that radius is 8.9.

In a forthcoming paper, in order to reconstruct the history of the AM\,1219-430 system and predict the 
evolution of the encounter, we will model the interaction between AM\,1219A and AM\,1219B through 
numerical simulations using the $N$-body/SPH code { \sc GADGET-2} \citep{springel05}. The model parameters for both galaxies will be constrained by the photometric parameters and kinematic properties derived in this work.

\end{enumerate}

\section*{Acknowledgements}

We thank anonymous referee for important
comments and suggestions that helped to
improve the contents of this manuscript. 
This work is based on observations obtained at the Gemini Observatory,
which is operated by the Association of Universities for Research
in Astronomy, Inc., under a cooperative agreement with the NSF
on behalf of the Gemini partnership: the National Science Foundation (United States), 
the National Research Council (Canada), CONICYT (Chile), the Australian Research Council 
(Australia), Minist\'erio da Ciencia e Tecnologia (Brazil) and SECYT (Argentina). 
This work has been partially supported by the Brazilian institutions Conselho 
Nacional de Desenvolvimento Cient\'ifico e Tecnol\'ogico (CNPq) and 
Coordena\c c\~ao de Aperfei\c coamento de Pessoal de N\'ivel Superior (CAPES).

\end{document}